\documentstyle[aps,twocolumn,prb,epsfig]{revtex}

\begin{document}

\newcommand{\hitc}{$\mathrm{high}-T_c$} 
\newcommand{\lotc}{$\mathrm{low}-T_c$}  
\newcommand{\etal}{\textit{et al.}}

\title{Mutual Inductance Route to Paramagnetic Meissner 
Effect in 2D Josephson-
Junction Arrays.}
\author{Cinzia De Leo and Giacomo Rotoli\dag}
\address{Dipartimento di Energetica and \dag Unita' INFM, Universita' 
dell'Aquila,\\
Localita' Monteluco, Roio Poggio I67040, L'Aquila, Italy.}
\author{Paola Barbara}
\address{Department of Physics, Georgetown University, Washington DC 20057.}
\author{A. P. Nielsen and C. J. Lobb}
\address{Center for Superconductivity Research,\\
University of Maryland, College Park MD 20742, USA.}
\date{\today}

\twocolumn[\hsize\textwidth\columnwidth\hsize\csname@twocolumnfalse\endcsname

\maketitle

\begin{abstract}
We simulate two-dimensional Josephson junction arrays, including full 
mutual-inductance effects, as they are cooled below the transition temperature in a 
magnetic field. We show numerical simulations of the array magnetization as a 
function of position, as detected by a scanning SQUID which is placed at a fixed 
height above the array. The calculated  magnetization images show striking 
agreement with the experimental images obtained by A. Nielsen \etal\cite{NIE}
The average array magnetization is found to be paramagnetic for many values of 
the applied field, confirming that paramagnetism can arise from magnetic screening 
in multiply-connected superconductors without the presence of $d$-wave 
superconductivity.
\end{abstract}


]

A DC paramagnetic susceptibility, reported first by Braunisch \etal\cite{BRAU} for
BSCCO, occurs in many \hitc\ superconductors when cooled through their transition
temperature in an external magnetic field.  
This surprising result, known as the paramagnetic Meissner effect (PME),
contrasts with the standard diamagnetic response of classical superconductors and has been subject of extensive investigations for the last ten years. 

Some theoretical work\cite{SIGR} suggested that the PME provided indirect 
evidence for $d$-wave symmetry in the superconducting order parameter. In this 
picture, $\pi$-junctions formed between misaligned grains were the cause of the 
anomalous magnetic response. 

PME
observed in \lotc\ superconductors with $s$-wave order parameters\cite{THOM}
demonstrated that $\pi$-junctions were not required for PME. 
New theories for
PME were developed, advocating non-equilibrium phenomena such as flux
compression,\cite{FLXC}
surface barriers \cite{SURF} and a giant vortex state.\cite{GVXS} However,
in the case of \hitc\ samples like BSCCO, experiments \cite{BRAU} showed 
clearly that the granular nature of the samples was a crucial ingredient 
for the occurrence of the phenomenon.
This suggested using arrays of (non-$\pi$) Josephson junctions\cite{NEWROCK} 
as a model system  for studying PME in granular \hitc\ samples, to test 
whether $\pi$-junctions were also an essential ingredient. Numerical
simulations of simplified Josephson junction networks (a single multi-junction 
loop\cite{pasqualina} or multi-junction concentric loops \cite{auletta}) 
indeed showed a paramagnetic response. Experiments also gave indirect 
evidence for PME in the AC susceptibility of arrays.\cite{MOREIRA}

Because of the many theories predicting PME in both $s$ and $d$-wave 
superconductors, more stringent and detailed experimental tests were 
needed to find the end of this maze. Experiments using scanning SQUIDs 
were thus performed on \hitc\ superconductors\cite{KIRTLEY} and on 
arrays of non-$\pi$ junctions.\cite{NIE} A scanning SQUID microscope 
(SSM) \cite{BLACK} measures the spatial distribution of the magnetization. The 
complexity of the results and the experimental technique pose new theoretical 
challenges in the qualitative and quantitative interpretation of the magnetic images. 

Here we show that a model of 2D arrays with full mutual inductance 
interactions captures the essential facts about PME\ in Josephson junction arrays.  

\begin{figure}[th]
\epsfig{file=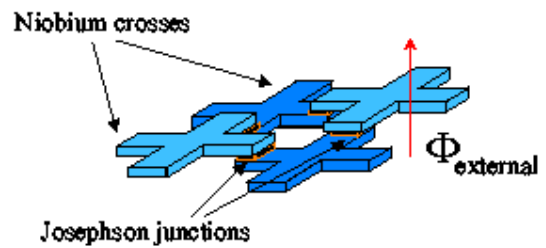,width=3.0in}
 \caption{Sketch of array design. Niobium crosses are in two layers, light
and dark grey. The Josephson junctions are formed at the cross overlaps,
as indicated, and the external flux is applied perpendicular to the 
array.}
 \label{design}
 \end{figure}

The arrays measured in Nielsen \etal\ had unit a cell size of 
$46\,\mu\mathrm{m}$
and were cooled in external flux from zero up to $12\,\Phi_0$ 
per unit cell of the 
array. A sketch of the array is shown in Fig.~1.
The junctions had a $J_c= 600\,\mathrm{A}/\mathrm{cm}^2$ with a 
junction area of 
$5\times 5\,\mu\mathrm{m}^2$ and a 
calculated self-inductance of $L' = 64\,\mathrm{pH}$, yielding a $\beta_L =
2 \pi L' I_0(T) / \Phi_0 = 30$ at $4.2\,\mathrm{K}$. The experiment involved 
cooling the array in an externally applied field and then measuring the magnetization 
with the field still applied. These parameters are similar to those in BSCCO 
which exhibits 
PME\cite{KIRTLEY} and are the parameters used here. 

We simulate a network of $N_{r}\times N_{c}$ junctions.%
\cite{ORPH,DOMJ,LUCH,EPJ} Using a vector notation,\cite{LUCH} the current in 
each junction can be modeled by the resistively-capacitively shunted 
junction (RCSJ)\ model as:
\begin{equation}
\vec{I}^{b}=I_{0}\sin \vec{\varphi}+
           {\Phi_0 \over 2\pi} G\overrightarrow{\dot{\varphi}}+
           {\Phi_0 \over 2\pi} C \overrightarrow{\ddot{\varphi}}.  
\label{rsj}
\end{equation}
Here $I_{0}\sin \vec{\varphi}$ represents the current through the
Josephson element ($\sin \vec{\varphi}$ is the vector given by applying $%
\sin $ to the components of $\vec{\varphi})$, and 
$(\Phi_0/2\pi) \overrightarrow{\dot{\varphi}}$
is the voltage drop across the quasiparticle conductance $G$.
Finally $C$ and $I_{0}$ are the junction
capacitance and the Josephson critical current.

To satisfy the Kirchhoff's law for the currents in each node, we define
loop currents
$\vec{I^{s}}$ connected to the junction currents by the
relationship $\vec{I^{b}}=\hat{K}\vec{I^{s}}$
(for a discussion of the role of loop currents cf. Refs. \onlinecite
{DOMJ,EPJ}) where the matrix $\hat{K}$ 
depends on the array geometry. The fluxoid quantization rule for
each elementary loop in the array gives another set of equations:
\begin{equation}
\hat{M}\vec{\varphi}=2\pi \vec{n}-2\pi \vec{f}+\frac{2\pi L^{\prime }}{\Phi
_{0}}\hat{L}\vec{I}^{s}  \label{fluxoid}
\end{equation}
where $\hat{M}$ performs the (oriented) sum of the phases around a
loop; the vector $\vec{f}$ represents the normalized flux $\phi
^{ext}=f\Phi _{0}$ due to an external field in each loop, i.e., the so-called frustration; $\vec{n}$ 
is a vector of ''quantum numbers'' for the flux quanta in each
loop; and the last term is the field induced by the currents flowing in all
other loops of the array ($\phi ^{induced}=L^{\prime }\hat{L}\vec{I}^{s}$). The
matrix $\hat{L}$, the mutual inductance matrix of the array
(normalized to the self inductance of the single loop),
represents the mutual coupling between loops in the arrays. Here we compute 
$\hat{L}$ by a thin wire approximation except for the self-inductance
of a single loop(cf. Ref.~\onlinecite{EPJ}). 
Inserting the fluxoid quantization in $\vec{%
I^{b}}=\hat{K}\vec{I^{s}}$, using Eq.(\ref{rsj}) we obtain  a system 
of equations in
normalized units, containing only the phase variables: 
\begin{equation}
\frac{\beta _{L}}{2\pi }\sin \vec{\varphi}+\sqrt{\frac{\beta _{L}}{\beta _{C}%
}}\overrightarrow{\dot{\varphi}}.+\overrightarrow{\ddot{\varphi}}=\hat{K}%
\hat{L}^{-1}\vec{m} .  \label{n-arr-eq}
\end{equation}
Here time is normalized to a cell frequency 
($\omega^{-2} ={L^{\prime} C}$) and
the usual Stewart-McCumber parameter appears,
$\beta_C = 2 \pi I_0(T) C / \Phi_0 G^2$.
The term $\vec{m}$ represents the normalized loop
magnetization (cf. Eq.~(\ref{fluxoid})). An explicit form for magnetization
can be written as follows by inverting the static form of Eq.(\ref
{n-arr-eq}):

\begin{equation}
\vec{m}=\frac{\beta _{L}}{2\pi }\hat{L}\left( \hat{K}^{T}\hat{K}\right) ^{-1}%
\hat{K}^{T}\sin \vec{\varphi},  \label{magn}
\end{equation}
which generalizes the single-loop Eq.(1) by Nielsen \etal\
In the case of a single loop, for large $\beta_L$, there are  at least four states 
which are non-degenerate and that are either
diamagnetic or paramagnetic. The lowest
energy states are  diamagnetic for $%
\ell<f<\ell+1/2$ with $\ell$ integer, and paramagnetic for $\ell+1/2<f<\ell+1$. For a
single loop, half the states are diamagnetic and half are paramagnetic.
This contrasts with the experiments on large arrays by Nielsen \etal\ that show a
clear prevalence of paramagnetism for $f\gtrsim 3$. In other words, the single-
loop model cannot explain the experimental results, even qualitatively.

\begin{figure}[th]
\epsfig{file=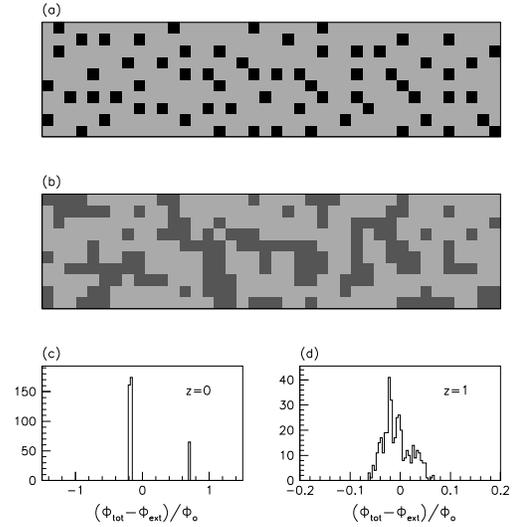,width=3.0in}
 \caption{Simulated field cooled 10 by 40 Josephson junction array  for a
frustration $f=1.2$. Parameters of simulations are $\beta _{L} (4.2~K)=30$, $\beta
_{C}(4.2~K)=66$. a) Image of the array magnetization at $z=0$;\ b)\ 
simulated SSM image of array
magnetization at $z=1$, sampled at positions corresponding to the center of array 
loops. The light-gray loops are the diamagnetic ones. c)\
Histogram of loop magnetization at $z=0$ ;\ d)\ histogram of
magnetization as read by SSM at $z=1$ .}
 \label{f1.2}
 \end{figure}

We can do a mean-field type of treatment of the temperature dependence by using 
the fact that $\beta_C$ and $\beta_L$ are the only temperature dependent quantities 
in these equations.  Thus, we simulated field cooling in the arrays solving Eq. 
(\ref{n-arr-eq})
for the phases and calculating the resulting currents and magnetization. The
simulation starts with a zero screening term in the equation, 
$\beta_{L}=0$, and $\beta_C=0$, as representing $T \ge T_c$.  Non-zero 
frustration $f$ was fixed in the beginning of the simulation. Then, $\beta _{L}$ 
and $\beta_C$ are increased in steps, until they reach their final, low-temperature 
value. The dynamical terms, i.e., $\ddot{\varphi}$ and $\dot{\varphi}$ go to 
zero after a transient. A variable transient time permits control of the speed 
of the simulated field cooling process. 
We used parameters similar to the experiments,\cite{NIE} i.e., $\beta _{L}(T = 
4.2\,\mathrm{K})=30$, $\beta _{C}(T = 4.2\,\mathrm{K})=66$. 
The transient time for each step 
increase in $\beta _{L}$  ranges from $80$ to $400$ normalized time units, and a 
typical run takes $30$ steps. The initial conditions for the array are chosen 
with all the phases being zero and a random distribution of ''quantum numbers'' 
$\vec{n}$, simulating the disorder due to the initial diffusion of flux quanta, 
when the Josephson energy barriers are small. Details of the integration routine are 
described in Filatrella \etal\cite{EPJ}

\begin{figure}[th]
\epsfig{file=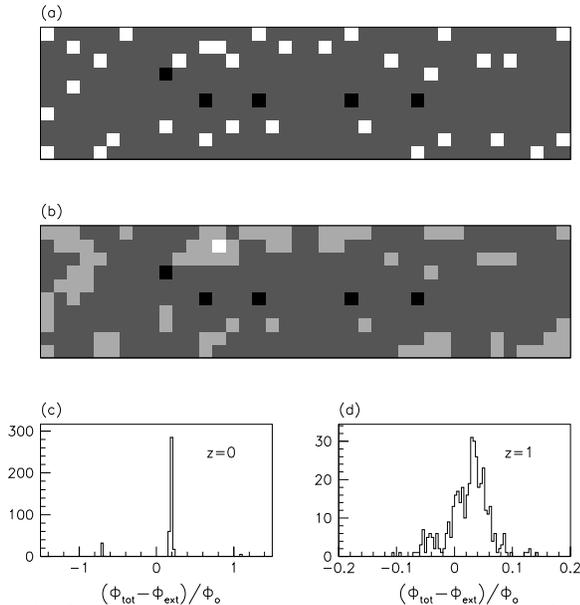,width=3.0in}
 \caption{The same simulated field cooled array of Fig.~2 for $f=4.8$. a) Image of the
 array magnetization at $z=0$;\ b)\ simulated SSM image of array magnetization at $z=1$, 
sampled at positions corresponding to the center of array loops. The light-gray loops are the diamagnetic ones. c)\ Histogram of
loop magnetization  at $z=0$ ;\ d)\ histogram of
magnetization values as read by SSM at $z=1$.}
 \label{f4.8}
 \end{figure}

In order to have a significant comparison between the numerical simulations and 
the experiments, we take into account the SQUID--sample separation 
at a non-zero distance $z$ above the array. Typical values of $z$ have 
been
chosen within the limits indicated by Ref.~\onlinecite{NIE}, $40$ to $60$ 
$\mu\mathrm{m}$, and we normalized $z$ to the array unit cell size, 
$46\,\mu\mathrm{m}$.
The field at a distance $z$ was built by
superposition of the fields generated by the currents. Each current in the array is modeled using the thin-wire approximation.\cite{ISEC2001}

Next, the flux within a square corresponding to the
SQUID\ area was calculated, for different positions above the array. We chose to 
calculate positions corresponding to the centers of the array loops (i.e. one 
point per loop) at distance $z$ above them.

Fig. 2 reports the field-cooled magnetization for a $10\times 40$
array with $f=1.2$ and clearly shows a diamagnetic behavior 
both
locally and in the average magnetization. Figs. 2a and 2b respectively show 
the magnetization 
at $z=0$ and $z=1$. Figs. 3 and 4 show the same
array for $f=4.8$ and $f=12.2$: Figs. 3a and 4a report the
magnetization at $z=0$, Figs.~3b and 4b magnetization at $z=1$. 
For values of 
frustration above 3, the array shows an overall paramagnetic response. 
It is interesting to note that in all cases, at $z=0$, there is a connection 
between the simulated arrays and the simple single loop picture. If, for a given 
value of frustration, an isolated loop is diamagnetic (lowest energy state), for 
the same value of frustration the simulated array shows a
larger number of diamagnetic loops. These diamagnetic loops form a ''sea'' in 
which a few paramagnetic loops stand out (cf. Figs.~2a, 4a).  If 
the isolated loop is paramagnetic, the ''sea'' is formed by paramagnetic loops 
with few diamagnetic loops in the array (cf. Fig.~3a).

\begin{figure}[th]
\epsfig{file=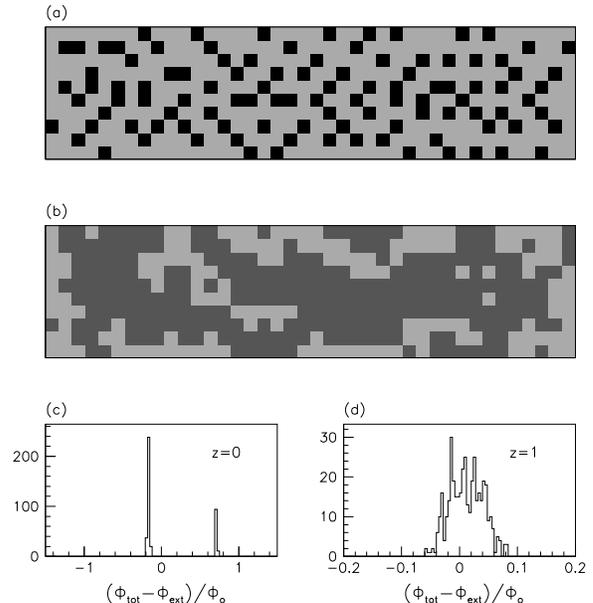,width=3.0in}
 \caption{The same simulated field cooled array of Fig.~2 for $f=12.2$. a) Image of the 
array magnetization at $z=0$;\ b)\ simulated SSM image of array magnetization at $z=1$, 
sampled at positions corresponding to the center of array loops. The light-gray meshes are the diamagnetic ones. c)\ Histogram of
loop magnetization at $z=0$ ;\ d)\ histogram of
magnetization values as read by SSM at $z=1$.}
 \label{f12.2}
 \end{figure}

At  $z=1$ the mixing of flux lines produces a 
smeared flux distribution that is 
very similar to the experiments (cf. Ref.~\onlinecite{NIE}). We note that for large 
frustration values (cf. Fig.s~4a and 4b), due to different
magnetization strength, the far-field array image is paramagnetic, although the
corresponding state for an isolated loop is diamagnetic.

In Figs.~2c, 3c and 4c, histograms of the loop magnetization are reported at $z=0$.
We find two peaks representing the diamagnetic ($\Phi_{tot}-\Phi_{ext}<0$) 
and paramagnetic ($\Phi_{tot}-\Phi_{ext}>0$) loops. 
The peak position essentially corresponds to single loop values for the same 
frustration. The peak width is determined by mutual inductance effects. 
Generally only two magnetization peaks are found, one diamagnetic and one paramagnetic (with 
the exception of a few loops in the $f=4.8$ case, which show a higher value of 
paramagnetic magnetization, cf. Fig.~3).
The majority loops magnetize weakly whereas the minority loops magnetize strongly. 
Figs.~2d, 3d, and 4d show the histograms evaluated at $z=1$.
Similarly to the measured images, we observe a smearing effect: Histogram peaks merge, so that the overall distributions appear similar to the experimental ones. Merging of histogram peaks starts approximatively at $z\simeq 0.3$. The results discussed for the Figs.~2, 3 and 4 can be extended to other frustration values:\cite{ToBePub}
Simulations show that for $\ell<f<\ell+1/2$, with $\ell$ integer, the diamagnetic loops predominate in number, whereas for $\ell+1/2<f<\ell+1$ the paramagnetic loops dominate. For $f=\ell+1/2$ the solution tends to have an equal number of diamagnetic and paramagnetic loops. The magnetization strength shows a more  subtle behavior:
For $\ell<f<\ell+1/2$ the strongest magnetization is paramagnetic, for $\ell+1/2<f<\ell+1
$ the strongest magnetization is diamagnetic. If the frustration equals 
a half integer, $f=\ell+1/2$ the paramagnetic and diamagnetic peaks are of equal strength, so their average magnetization measures zero. 

\begin{figure}[t]
\epsfig{file=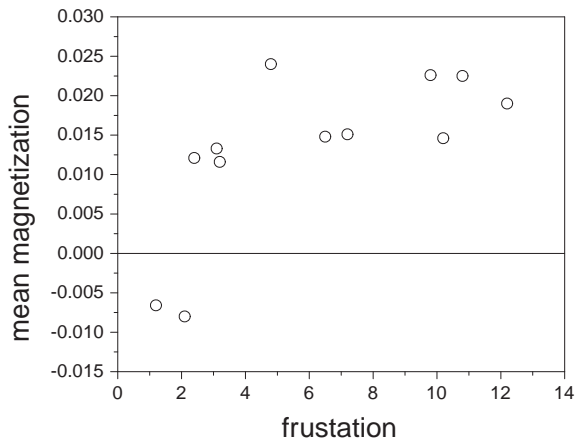,width=3.0in}
 \caption{Dependence of mean array magnetization on frustration for a 10 by 40
array. Parameters of simulations are $\beta _{L}(4.2~K)=30$, $\beta _{C}(4.2~K)=66$, $z=1$.}
 \label{f-all}
 \end{figure}

In Fig.~5 we report the mean magnetization over a $10\times 40$ array, for 
different frustrations, at $z=1$. The mean magnetization depends on the blend
of paramagnetic/diamagnetic strength in the loops and their number. A 
trend shifts the array magnetization toward paramagnetism, starting from 
$f\gtrsim 3$.
The mean magnetization depends weakly on noise: A test with different
random distributions of quantum numbers shows that this accounts for an error of 
about $1.2\%$. Our estimation of $z$ adds another source of error, but
our simulations show that 
this error accounts for no more than $5\%$, with a $z$ variance of $20\%$. 
On the
other hand, the mean magnetization depends on the dimension of the array. A 
direct quantitative comparison with the experiments shows a calculated value of magnetization typically lower. 
Magnetization strongly depends on array dimension,
so the results presented in Fig.~5
can be only
qualitatively compared with experiments, in which arrays are larger. 
We report only positive frustration ($f>0$) because Eq.(\ref{n-arr-eq}) 
is symmetric changing the sign of frustration (the same array viewed from below 
simply maintains the same para and diamagnetic loops).

We note that in all cases, i.e., both dia- and paramagnetic, 
diamagnetic behavior prevails near the array edges. This agrees 
with the experiments, which show a similar behavior. According to 
Ref.~\onlinecite{NIE} 
this occurs because the array screens the field by generating diamagnetic 
currents on the array boundary and, as consequence, induces paramagnetic 
currents in the interior of the array, thus generating an overall paramagnetic 
offset.

To further support this view, we calculated the densities of paramagnetic loops 
at the boundary and in the bulk of the array. We find that
there is a clear divergence between two data sets with an increase of bulk
density, $\rho_{k}$ with respect to boundary density, $\rho_{b}$, for frustration 
$f\gtrsim 3$.
 For example, at $f=1.2$ the two densities are roughly equal $\rho%
_{b}=N_{para}/N_{boundary}\simeq 0.156$ and $\rho%
_{k}=N_{para}/N_{total}\simeq 0.162$, but at $f=12.2$ at the boundary we
have $\rho _{b}\simeq 0.11$ and in the bulk $\rho_{k}\simeq 0.26$. Tests on
smaller arrays show that paramagnetic behavior for $m<f<m+1/2$ arise about
for $N\sim 5$, this is roughly the value predicted from Eq.~(4)\ of 
Ref.~\onlinecite{NIE}
for $\beta _{L}=30$.

In conclusion, the PME in Josephson junction arrays can be reproduced via 
numerical simulations which include the full inductance matrix. 
The simulation results compare favorably to experimental results: Paramagnetism dominates field cooling for large arrays. 
Simulations also show that the single loop model is the basic building block describing the field cooled array behavior. Mutual inductance interactions create
the actual distribution of loop magnetization in the arrays. The resulting mean 
magnetization is the product of both single loop states and their occupancy.
The observed dominant paramagnetism, 
in both experiments and simulations, arises from an energetic preference for 
paramagnetic loops interior to the array. 

Beyond this study, a number of open problems still remain to be analyzed.
Among these, are simulations of larger arrays in order to make more detailed 
comparison with experiments and the study of the effect of cooling time and 
transient dynamics of the array.

We acknowledge support by MURST\ COFIN98 project ''Dynamics and
Thermodynamics of vortex structures in superconductive tunneling,'' 
by AFOSR under grant F4620-98-1-0072 and by NSF under grant DMR 9732800.

 \end{document}